\documentclass[cits]{PoS}
\usepackage{amssymb,amsmath}

\newcommand{\lsim}{
\mathrel{\hbox{\rlap{\hbox{\lower4pt\hbox{$\sim$}}}\hbox{$<$}}}}

\newcommand{\gsim}{
\mathrel{\hbox{\rlap{\hbox{\lower4pt\hbox{$\sim$}}}\hbox{$>$}}}}

\title{\boldmath $K$ and $B$ Physics in the Custodially Protected Randall-Sundrum Model}

\ShortTitle{$K$ and $B$ Physics in the Custodially Protected Randall-Sundrum Model}

\author{\speaker{Monika Blanke}
\\
        Physik Department, Technische Universit\"at M\"unchen,
D-85748 Garching, Germany\\
        E-mail: \email{mblanke@ph.tum.de}}

\abstract{The Randall-Sundrum framework provides an appealing solution to both the gauge hierarchy and the flavour problem. The model with enlarged custodial symmetry is consistent with electroweak precision observables for Kaluza-Klein (KK) scales as low as $(2-3)\,\text{TeV}$, so that the lowest-lying KK modes could be observed at the LHC. The presence of the heavy KK gauge bosons induces new tree level contributions to flavour violating observables. We summarise the main results for $K$ and $B$ physics observables in the $\Delta F=2$ and $\Delta F=1$ sectors. The specific pattern of new physics effects can be used to distinguish this model from other new physics frameworks.}

\FullConference{The 2009 Europhysics Conference on High Energy Physics,\\
		 July 16 - 22 2009\\
		 Krakow, Poland}

\begin{document}

\section{The custodially protected Randall-Sundrum model}

During the past ten years Randall-Sundrum (RS) models~\cite{Randall:1999ee} have attracted a lot of attention. Apart from their original motivation to solve the gauge hierarchy problem, models with bulk fields~\cite{Chang:1999nh} are also able to address the flavour problem~\cite{Huber:2003tu,Agashe:2004cp}. As the localisation of the fermionic zero modes depends exponentially on their bulk mass parameters, the observed hierarchies in the effective Yukawa couplings can naturally be generated from $\mathcal{O}(1)$ fundamental parameters. 

The simplest RS model with only the SM gauge group in the bulk has severe problems with the electroweak precision parameters, so that the lowest-lying KK modes have to be at least $\sim10\,\text{TeV}$ and therefore beyond the LHC reach. However RS models with an enlarged custodial symmetry~\cite{Agashe:2003zs} can be made consistent with electroweak precision data for KK scales even as low as $(2-3)\,\text{TeV}$~\cite{Carena:2007ua}. A detailed description, including a set of Feynman rules, of such a custodially protected RS model has been presented in~\cite{Albrecht:2009xr}, where further references can be found.

As the fermions' couplings to the KK gauge bosons depend on the overlaps of the respective bulk profiles, flavour non-universalities and therefore flavour changing neutral currents (FCNCs) arise already at the tree level. These effects are transmitted also to the $Z$ couplings via electroweak symmetry breaking. Fortunately the protective RS-GIM mechanism \cite{Agashe:2004cp} is at work: FCNCs are strongly suppressed by the hierarchies responsible for the effective Yukawa couplings. Still interesting and potentially large new physics (NP) contributions to flavour violating observables arise.


\section{Meson-antimeson mixing and fine-tuning \cite{Blanke:2008zb}}

\begin{figure}[b]
\center{\begin{minipage}{6cm}
\includegraphics[width=\textwidth]{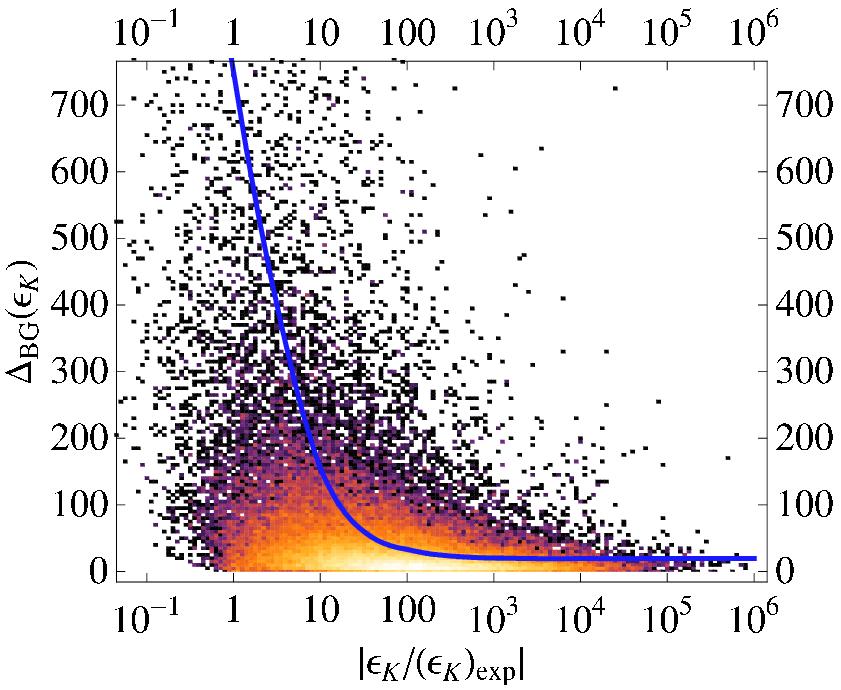}
\end{minipage}
\hspace{1cm}
\begin{minipage}{6cm}
\includegraphics[width=\textwidth]{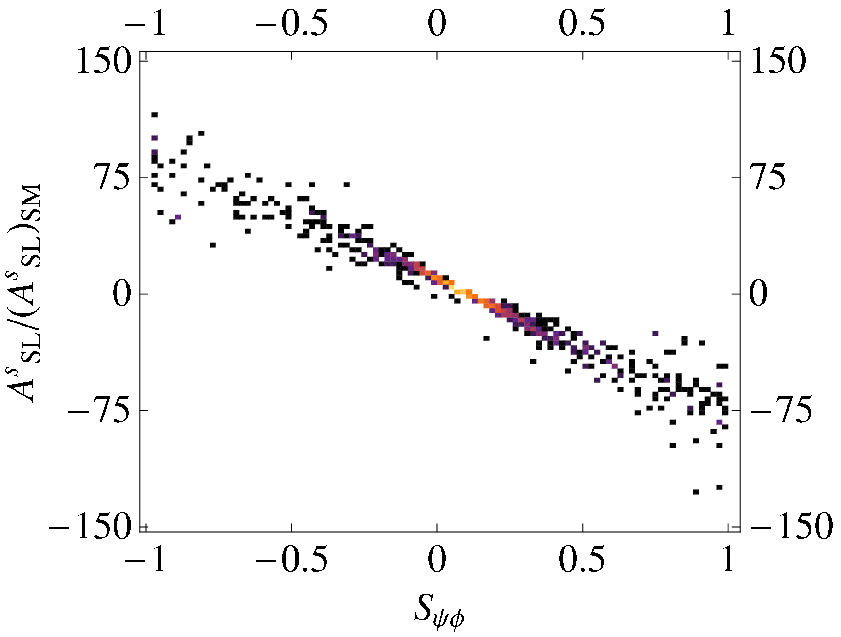}
\end{minipage}}
\hfill
\caption{Left: Fine-tuning $\Delta_\text{BG}(\varepsilon_K)$ as a function of $\varepsilon_K$. Right: Correlation between $S_{\psi\phi}$ and $A^s_\text{SL}$. \cite{Blanke:2008zb}}
\label{fig1}
\end{figure}

Meson-antimeson mixings in the $K$ and $B_{d,s}$ systems receive significant contributions from KK gluons and the new heavy $Z_H$ gauge boson. In particular new contributions to the scalar left-right operator $(\bar s P_L d)(\bar s P_Rd)$ are generated by the exchange of KK gluons which, due to the strong chiral and QCD enhancement, lead to large NP effects in $K-\bar K$ mixing. Assuming anarchic, i.\,e.\ structureless 5D Yukawa couplings, the constraint from the CP-violating parameter $\varepsilon_K$ leads to the generic lower bound $M_\text{KK}\gsim 20\,\text{TeV}$ \cite{Csaki:2008zd}.
However, by partly abandoning the anarchic ansatz, it is possible to lower the KK scale down to $M_\text{KK}\simeq 2.5\,\text{TeV}$. As we can see in the left panel of Fig.\,\ref{fig1} even for such low KK masses it is possible to obtain an agreement with the $\varepsilon_K$ data without the need for large fine-tuning in the fundamental 5D Yukawa couplings.

The data on other $\Delta F=2$ observables turn out to be less restrictive, so that a simultaneous agreement with all these observables can be obtained. Having imposed these constraints we show in the right panel of Fig.\,\ref{fig1} the correlation between the mixing-induced CP-asymmetry $S_{\psi\phi}$ and the semi-leptonic CP-asymmetry $A^s_\text{SL}$, both measuring CP-violating effects in $B_s-\bar B_s$ mixing. Thanks to the new sources of flavour and CP-violation, the custodially protected RS model can easily account for the recently observed possible non-SM effects in that system \cite{Punzi}.

\section{\boldmath Rare $K$ and $B$ decays~\cite{Blanke:2008yr}}

\begin{figure}
\center{\begin{minipage}{6.4cm}
\includegraphics[width=\textwidth]{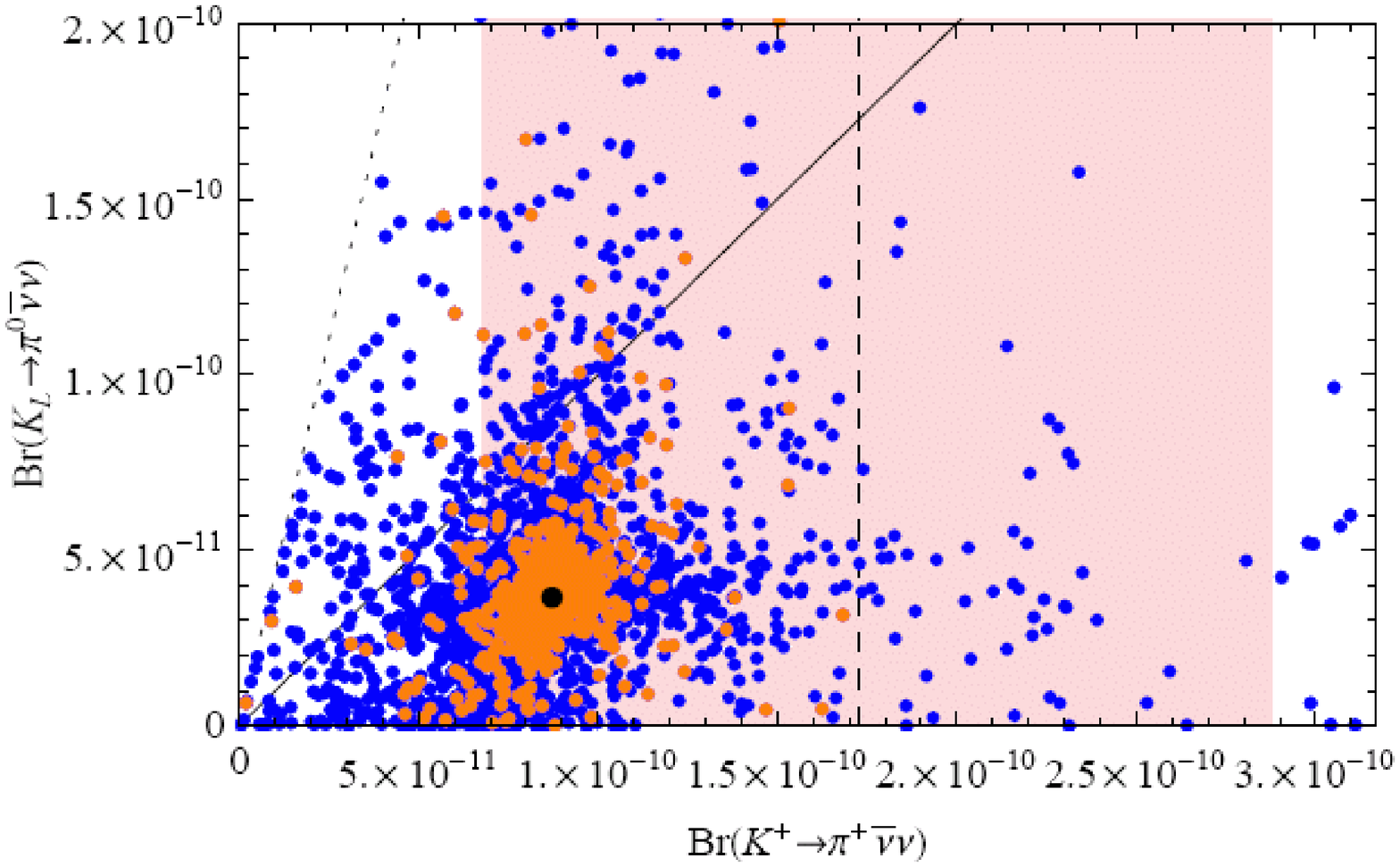}
\end{minipage}
\hspace{1cm}
\begin{minipage}{6.4cm}
\includegraphics[width=\textwidth]{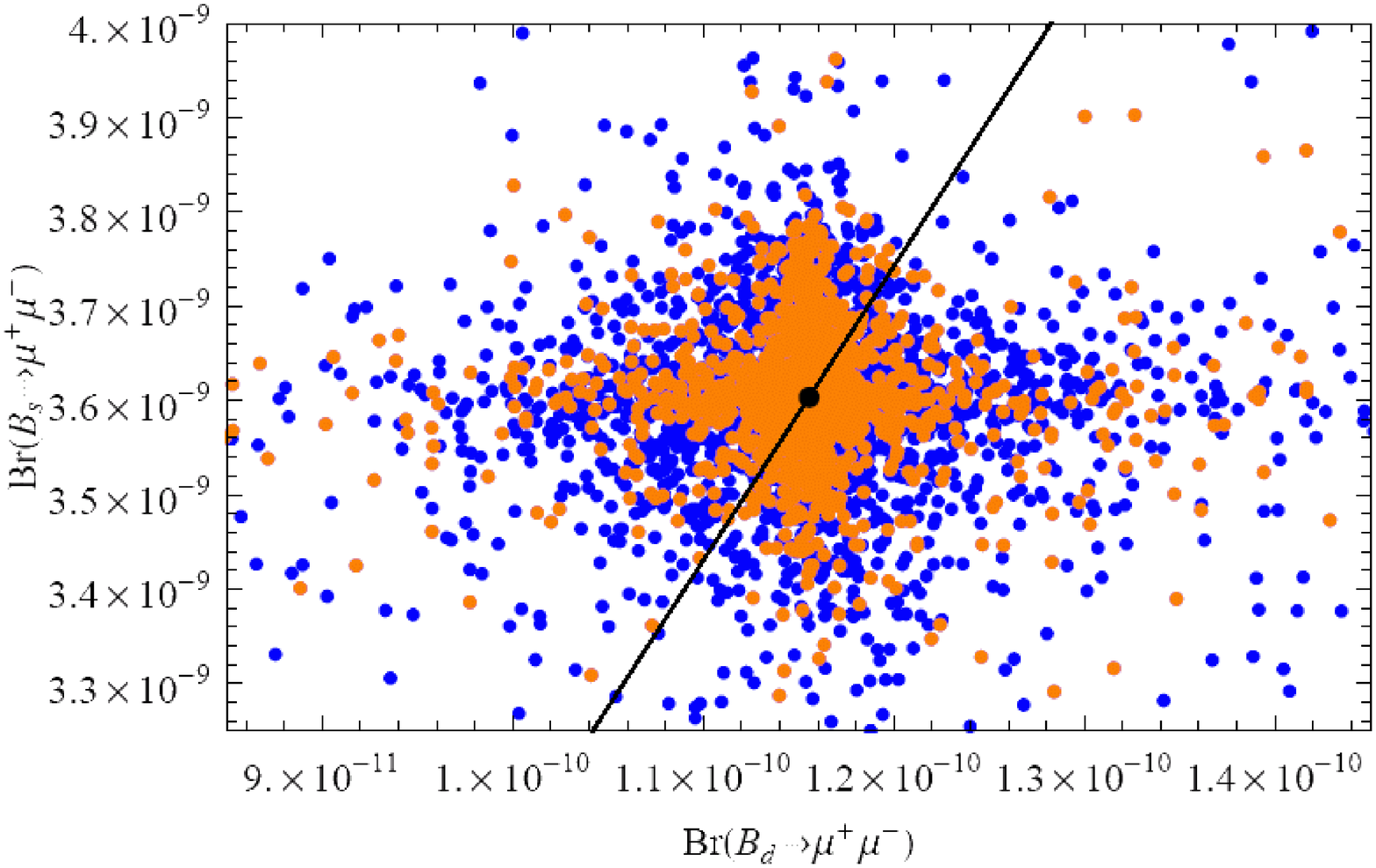}
\end{minipage}}
\caption{Left: The $K\to\pi\nu\bar\nu$ plane. Right: $Br(B_s\to\mu^+\mu^-)$ and $Br(B_d\to\mu^+\mu^-)$. \cite{Blanke:2008yr}}
\label{fig2}
\end{figure}

In contrast to the $\Delta F=2$ sector where the dominant NP contributions arise from KK gauge boson exchanges, $\Delta F=1$ processes are dominated by the new tree level contributions of the $Z$ boson.
As the left-handed $Z d_L^i \bar d_L^j$ couplings are protected by the enlarged custodial symmetry~\cite{Blanke:2008zb,Buras:2009ka}, rare $K$ and $B$ decays are dominated by the $Z$ boson couplings to right-handed down-type quarks. Consequently, as the hierarchy in the right-handed down sector can only partly compensate the CKM hierarchy, the NP effects in $K$ decays are much larger than in $B$ decays.

As seen from the left panel of Fig.\,\ref{fig2}, the $K\to\pi\nu\bar\nu$ branching ratios can be enhanced by as much as $(100-200)\%$ with respect to the SM. In addition, as a result of the new left-right operator contributions to $\varepsilon_K$ no visible correlation arises \cite{Blanke:2009pq}. Thus the $K\to\pi\nu\bar\nu$ decays can be used to distinguish this NP model from other frameworks such as the Littlest Higgs model with T-parity whose specific flavour structure results in a strict two-branch correlation in the $K\to\pi\nu\bar\nu$ plane \cite{LHT}.
Also in the case of other rare $K$ decays large deviations from the SM can be found and specific correlations arise that can be used to test this model. Interestingly large effects in the $K$ decays can {\em not} appear simultaneously with a large non-SM $S_{\psi\phi}$.

On the other hand the NP effects in rare $B$ decays are generally small ($\mathcal{O}(10\%)$) in the model in question and therefore difficult to disentangle from the SM (see e.\,g.\ the $B_{s,d}\to\mu^+\mu^-$ modes in the right panel of Fig.\,\ref{fig2}). Due to the presence of new flavour and CP-violating interactions the predictions from models with Minimal Flavour Violation can be strongly violated.

The analysis of correlations among various flavour violating observables thus provides a promising tool to understand the origin of NP, complementary to direct searches at collider experiments.

\section*{Acknowledgements}

Warmest thanks are given to my collaborators in the project summarised here: M.\,E.\,Albrecht, A.\,J.\,Buras, B.\,Duling, K.\,Gemmler, S.\,Gori and A.\,Weiler.  This work was partially supported by the Max-Planck-Institut f\"ur Physik M\"unchen, the Deutsche Forschungsgemeinschaft (DFG) under contract BU 706/2-1 and the DFG Cluster of Excellence `Origin
and  Structure of the Universe'.

\end{document}